\documentclass[reprint, showpacs, superscriptaddress, aps, pre, twocolumn, 10pt]{revtex4-1}

\usepackage{graphicx} 
\usepackage{dcolumn}
\usepackage{bm}
\usepackage{amssymb}
\usepackage{amsmath}
\usepackage{wasysym}
\usepackage{color}
\usepackage{hyperref}
\usepackage{dcolumn}
\usepackage{float}
\usepackage{flushend}
\usepackage{soul}
\usepackage{enumerate}

\hypersetup{
colorlinks,
citecolor=blue,
filecolor=blue,
linkcolor=blue,
urlcolor=blue}

\begin{document}
\title{Energy Spectrum of Buoyancy-Driven Turbulence}
\author{Abhishek Kumar}
\affiliation {Department of Physics, Indian Institute of Technology Kanpur, Kanpur, India 208016}
\author{Anando G. Chatterjee}
\affiliation {Department of Physics, Indian Institute of Technology Kanpur, Kanpur, India 208016}
\author{Mahendra K. Verma}
\email{mkv@iitk.ac.in}
\affiliation {Department of Physics, Indian Institute of Technology Kanpur, Kanpur, India 208016}

\begin{abstract}
Using high-resolution direct numerical simulation and arguments based on the kinetic energy flux $\Pi_u$, we demonstrate  that for stably stratified flows, the kinetic energy spectrum $E_u(k) \sim k^{-11/5}$,  the entropy spectrum $E_\theta(k) \sim k^{-7/5}$, and  $\Pi_u(k) \sim k^{-4/5}$, consistent with the Bolgiano-Obukhov scaling. This scaling arises due to the conversion of kinetic energy to the potential energy by buoyancy.   For weaker buoyancy, this conversion is weak, hence $E_u(k)$ follows Kolmogorov's spectrum with a constant energy flux.  For Rayleigh B\'{e}nard convection, we show that the energy supply rate by buoyancy is positive, which leads to an increasing $\Pi_u(k)$ with $k$, thus ruling out Bolgiano-Obukhov scaling for the convective turbulence.  Our numerical results show that convective turbulence for unit Prandt number exhibits a constant $\Pi_u(k)$ and $E_u(k) \sim k^{-5/3}$ for a narrow band of wavenumbers.
\end{abstract}

\maketitle

\section{Introduction}
\label{sec:intro}

Buoyancy or  density gradients drive flows in the atmosphere and interiors of planets and stars, as well as in electronic devices and industrial appliances like heat exchangers, boilers, etc.  Accordingly, scientists (including geo-, astro-, atmospheric- and solar physicists) and engineers have been studying buoyancy-driven flows for more than a century.  An important unsolved problem in this field is how to quantify the spectra and fluxes of kinetic energy and entropy ($u^2/2$ and $\theta^2/2$ respectively, where $\mathbf u$ and $\theta$ are the velocity and temperature fluctuations) of buoyancy-driven flows~\cite{Siggia:ARFM1994,Lohse:ARFM2010}. In this paper, we will study these quantities and respective nonlinear fluxes using direct numerical simulations, and show that the kinetic energy (KE) spectrum differs from  Kolmogorov's theory when buoyancy is strong.

Flows driven by buoyancy can be classified in two categories: (a) convective flows in which hotter and lighter fluid at the bottom rises, while  colder and heavier fluid at the top comes down;  (b) stably stratified flows in which lighter fluid rests above  heavier fluid.     The convective flows are unstable; but the stably stratified flows are  stable, hence their fluctuations vanish over time. Therefore, a steady state of a stably stratified flow is achieved only when it is driven by an external force.  Even though both types of flows are driven by  density gradients, the properties of such flows are quite different, which we decipher using quantitative analysis of energy flux and energy  supply rate by buoyancy.

For a stably stratified flow, Bolgiano~\cite{Bolgiano:JGR1959} and Obukhov~\cite{Obukhov:DANS1959} first proposed a phenomenology, according to which the KE flux $\Pi_u$ of a stably stratified flow is depleted at different length scales due to a conversion of  KE to ``potential energy'' via buoyancy ($u_z \theta$).   As a result, $\Pi_u(k)$ decreases with wavenumber, and the energy spectrum is steeper than that prediced by Kolmogorov theory $(E(k) \sim k^{-5/3}$, where $k$ is the wavenumber);  we refer to the above as    {\em BO} phenomenology or scaling.  According to this phenomenology, for $k<k_B$, where $k_B$ is the Bolgiano wavenumber~\cite{Bolgiano:JGR1959}, the KE spectrum $E_u(k)$, entropy spectrum $E_\theta(k)$,  $\Pi_u$, and entropy flux $\Pi_\theta$ are:
\begin{eqnarray}
E_u(k) & =  & c_1 (\alpha^2 g ^2 \epsilon_\theta)^{2/5}k^{-11/5}, \label{eq:Eu} \\
E_\theta(k) & =  & c_2 (\alpha g)^{-2/5}\epsilon_\theta^{4/5} k^{-7/5}, \label{eq:Etheta} \\
\Pi_\theta(k) & = &  \epsilon_\theta = \mathrm{constant}, \label{eq:pi_theta} \\
\Pi_u(k) & = & c_3 (\alpha^2 g^2 \epsilon_\theta)^{3/5} k^{-4/5},  \label{eq:pi}
\end{eqnarray}
where $\alpha$, $g$, and $\epsilon_{\theta}$ are the thermal expansion coefficient, acceleration due to gravity, and the entropy dissipation rate respectively, and $c_i$'s are constants.  For the wavenumbers in the range $k_B < k < k_d$,  $E_u(k), E_\theta(k) \sim k^{-5/3}$, and $\Pi_u \approx \epsilon_u$, where $\epsilon_u$ is the KE dissipation rate, and $k_d$ is the wavenumber after which dissipation starts.   We remark that many researchers describe the stably stratified flows in terms of density fluctuation $\rho'$, which leads to an equivalent description since $\rho' \propto -\theta$. 

Several research groups studied  the properties of stably stratified flows using numerical simulations.  Kimura and Herring~\cite{Kimura:JFM1996}  observed {\em BO} scaling in a narrow band of wavenumbers in their $128^3$ decaying buoyancy-dominated simulation.  In 2012, using $1024^3$ simulations, Kimura and Herring~\cite{Kimura:JFM2012} showed that waves and vortex  exhibit $k^{-5/3}$ energy spectra at large wavenumbers, but for sufficiently strong stratification, the corresponding spectra are $k_\perp^{-2}$ and $k_\perp^{-3}$, respectively, at small wavenumbers.  

The  terrestrial atmosphere exhibits   $k^{-3}$ energy spectrum for $k < 1/500~\mathrm{km}^{-1}$, and  $k^{-5/3}$ spectrum  for $k > 1/500~\mathrm{km}^{-1}$.  Lindborg~\cite{ Lindborg:GRL2005,Lindborg:JFM2006} and Brethouwer et al.~\cite{Brethouwer:JFM2007} attempted to explain this observation by studying quasi two-dimensional stratified flow (horizontal distance $\gg$ vertical distance).  They performed a series of periodic box simulations and showed that the horizontal kinetic and potential energy spectra follow $k_{\perp}^{-5/3}$ scaling, while the kinetic energy spectrum of the vertical velocity, and the potential energy spectrum follow $k_{\parallel}^{-3}$ scaling. Vallgren et al.~\cite{Vallgren:PRL2011}, and Bartello and Tobias~\cite{Bartello:JFM2013} observed similar scaling in their numerical simulations.  It is important to note that all these work are under the regime of strong stratification.

Using theoretical arguments, Procaccia and Zeitak~\cite{Procaccia:PRL1989}, L'vov~\cite{Lvov:PRL1991},  L'vov and Falkovich~\cite{Lvov:PD1992}, and Rubinstein~\cite{Rubinstein:NASA1994} proposed that the {\em BO} scaling would also be  applicable to Rayleigh-B\'{e}nard convection (RBC).  The numerical and experimental results of RBC, however, have been largely inconclusive.   Based on simulations with periodic boundary conditions, Borue and Orszag~\cite{Borue:JSC1997} and \v{S}kandera et al.~\cite{Skandera:HPCISEG2SBH2009} reported Kolmogorov-Obukhov (referred to as  {\em KO}) scaling, in which $\Pi_u \approx  \mathrm{const}$, and $E_u(k), E_\theta(k) \sim k^{-5/3}$.    Mishra and Verma~\cite{Mishra:PRE2010} reported the {\em KO} scaling for zero- and low Prandtl number flows.  Using numerical simulations, Verzicco and Camussi~\cite{Verzicco:JFM2003,Camussi:EJMF2004} however reported the {\em BO} scaling for the frequency spectrum, which was computed using the data collected by real space probes.     Calzavarini et al.~\cite{Calzavarini:PRE2002} reported the {\em BO} scaling in the boundary layer, and the {\em KO} scaling in the bulk.     The experimental results~\cite{Wu:PRL1990,Chilla:INCD1993,Cioni:EPL1995,Niemela:NATURE2000,Zhou:PRL2001,Shang:PRE2001,Mashiko:PRE2004,Zhang:PRL2005,Sun:PRL2006} are  more divergent  with some reporting the {\em BO} scaling, and some others reporting the {\em KO} scaling.

In this paper we simulate the stably stratified and RBC turbulence, and  analyse the spectra and fluxes of the KE as well as the entropy.  We show that for the stratified flow, the KE flux and spectrum follow the {\em BO} scaling (Eqs.~(\ref{eq:Eu}-\ref{eq:pi})) when buoyancy is strong, but they follow the {\em KO} scaling for weak buoyancy.   The KE flux in RBC however  increases at small wavenumbers, but remains flat for a narrow wavenumber band in the intermediate regime where the energy spectrum follows the {\em KO} scaling.

The outline of the paper is as follows. In Sec.~\ref{sec:gov_eqns}, we describe the parameters and equations used, as well as our assumptions.  In Sec.~\ref{sec:method}, we discuss the numerical method of our simulations. Results of our numerical simulations are discussed in Sec.~\ref{sec:results}.   We present our conclusions in Sec.~\ref{sec:conclusion}.

\section{Energy flux and spectrum in buoyancy-driven flows}
\label{sec:gov_eqns}

\subsection{Governing equations and assumptions}
The dynamical equations that describe the buoyancy-driven flows under the Boussinesq approximation are
\begin{eqnarray}
\frac{\partial \bf u}{\partial t} + (\bf u \cdot \nabla) \bf u & = & -\frac{\nabla \sigma}{\rho_0} + \alpha g \theta \hat{z} + \nu \nabla^2 \bf u + \bf f^u, \label{eq:u_dim} \\
\frac{\partial \theta}{\partial t} + (\bf u \cdot \nabla) \theta & = & S  \frac{\Delta}{d} u_z + \kappa \nabla^2 \theta, \label{eq:th_dim} \\
\nabla \cdot \bf u & = & 0 \label{eq:inc_dim}, 
\end{eqnarray}
where ${\bf u}$ is the velocity field, $\theta$  and $\sigma$ are the temperature and pressure fluctuations, respectively, with reference to the conduction state,  $\hat{z}$ is the buoyancy direction, $\bf f^u$ is the external force field,  $\Delta$ is the temperature difference between  two layers kept apart by a vertical distance $d$, and $\rho_0$, $\nu$, and $\kappa$ are fluid's  mean density,  kinematic viscosity, and thermal diffusivity respectively.  For RBC, temperature of the top plate is lower than the bottom one, hence $S=+1$, but for the stably stratified flows, the gradient is opposite, i.e. $S=-1$.  

It is easy to verify that Eqs.~(\ref{eq:u_dim},\ref{eq:th_dim}) conserve the volume integral $\int (u^2 - S  \alpha g d \theta^2 /\Delta) d\mathbf x$ in limit when $\nu = \kappa = 0$ and ${\bf f}^u = 0$.  

For RBC, the temperature gradient provides energy to the system, and a steady state is reached after some time (approximately after a thermal diffusive time); for such flows we can take ${\bf f}^u=0$.  However, stably stratified flows are stable, and the fluctuation die out if ${\bf f}^u=0$.  Therefore, for obtaining a steady state in a stably stratified flow, we force the flow at small wavenumbers with random forcing prescribed by Kimura and Herring~\cite{Kimura:JFM2012}. 

In this paper, we contrast the  scaling relations of stably stratified flow and RBC in a single formalism.  For the same, we use temperature fluctuations $\theta$ as a variable.  However, this scheme is equivalent to usage of $\rho'$, the density fluctuations from the linear density profile $\bar{\rho}$;  the variable $\rho'$ is often used for stably stratified flows.  We can rewrite Eqs.~(\ref{eq:u_dim}-\ref{eq:inc_dim})  in terms of $\rho'$ using the following relations:
\begin{equation}
\frac{\rho'}{\rho_0} = -\alpha \theta; ~~~~ \frac{d \bar{\rho}}{dz} = -\frac{\rho_0 \alpha \Delta}{d},
\end{equation}
thus, the two sets of equations are equivalent. 

It is convenient to work with nondimensionalized equations, which is achieved by using $d$ as the length scale,  $\sqrt{\alpha g \Delta d }$ as the velocity scale, and $\Delta$ as the temperature scale. Therefore, ${\mathbf u} =   {\mathbf u}'\sqrt{\alpha g \Delta d }$, $\theta = \theta' \Delta$, $\mathbf x =  \mathbf x' d$, and $t= (d/\sqrt{\alpha g \Delta d })t'$, where primed variables are nondimensionalized.  When we use the density gradient $d\bar{\rho}/dz$, the velocity scale is  $d \sqrt{ g  (d \bar{\rho}/dz)/\rho_0 }$, and the time scale is $1/\sqrt{ g  (d \bar{\rho}/dz)/\rho_0 }$. 
In terms of the  nondimensionalized variables, the equations are
\begin{eqnarray}
\frac{\partial \bf u'}{\partial t'} + (\bf u' \cdot \nabla') \bf u' & = & -\nabla' \sigma' + \theta' \hat{z} + \sqrt{\frac{\mathrm{Pr}}{\mathrm{Ra}}} \nabla'^2 \bf u' + \bf f'^u, \label{eq:u_ndim} \\
\frac{\partial \theta'}{\partial t'} + (\bf u' \cdot \nabla') \theta' & = & S  u'_z + \frac{1}{\sqrt{\mathrm{Ra}\mathrm{Pr}}}\nabla'^2 \theta', \label{eq:th_ndim} \\
\nabla' \cdot \bf u' & = & 0 \label{eq:inc_ndim}, 
\end{eqnarray}
where  the Prandtl number is defined as
\begin{equation}
\mathrm{Pr} = \frac{\nu}{\kappa}
\end{equation} 
the Rayleigh number is defined as
\begin{equation}
\mathrm{Ra}_1 = \frac{\alpha g \Delta d^3}{\nu \kappa};~~ \mathrm{Ra}_2 = \frac{d^4 g}{\nu \kappa \rho_0}\left| \frac{d \bar{\rho}}{dz} \right| = \frac{N^2 d^4}{\nu \kappa},
\end{equation}
where $\mathrm{Ra}_1$ is the usual definition taken from RBC, but $\mathrm{Ra}_2$, a modified form of $\mathrm{Ra}_1$, is in terms of density gradient and   Brunt V\"ais\"al\"a frequency, which is defined as
\begin{equation}
N = \sqrt{ \frac{g}{\rho_0}  \left| \frac{d \bar{\rho}}{dz} \right|}.
\end{equation}
Physically, Brunt V\"ais\"al\"a frequency is the frequency of the gravity waves in a stably stratified flow.  It is important to note that larger $ \mathrm{Ra}_2$ or $N$ implies stronger {\em stability} for a stably stratified flow, but larger $\mathrm{Ra}_1$ implies stronger {\em instability} for RBC.  Also, it has been shown that  the ``available potential energy (APE)", $\int ( \rho' g z) d \bf x$, matches with $\int ( \rho_0 b'^2 /2) d \bf x$ where $b'= \rho' g / \rho_0 N$~\cite{Lorenz:Tellus1954,Davidson:book_2}.
 
 The other important nondimensional numbers are as follows. The Reynolds number is defined as
 \begin{eqnarray}
\mathrm{Re} & = & \frac{u_{\rm rms} d}{\nu} = \frac{u'_{\rm rms}  d^2 \sqrt{g  (d \bar{\rho}/dz)/\rho_0 }  }{\nu}  \\
 & = & \frac{u'_{\rm rms} N d^2}{\nu} = u'_{\rm rms}  \sqrt{\frac{\mathrm{Ra}}{\mathrm{Pr}}},
\label{eq:Re}
\end{eqnarray}
where $u_{\rm rms}$ is the rms velocity of the flow,  computed as the volume average of the magnitude of the velocity field, and $u'_{\rm rms}$ is the corresponding quantity in dimensionless form.  The Richardson number, which  is a ratio of the buoyancy and the nonlinear term $(\bf u \cdot \nabla) \bf u$,  is defined as
\begin{equation}
\mathrm{Ri} = \frac{\alpha g \Delta d }{u_{\rm rms}^2} = \frac{1}{u_{\rm rms}^{'2}}.
\label{eq:Ri}
\end{equation}
The Froude number $\mathrm{Fr}$, which is the ratio of the characteristic fluid velocity and gravitational wave velocity, is defined as
\begin{equation}
\mathrm{Fr} = \frac{u_\mathrm{rms}}{d N} = \frac{u'_\mathrm{rms} \sqrt{ g d^2 (d \bar{\rho}/dz)/\rho_0 }} {d \sqrt{(g/\rho_0) d \bar{\rho}/dz}} = u'_\mathrm{rms}.
\label{eq:Fr}
\end{equation}
Thus, the Froude number is the rms velocity of the fluid in the dimensionless form.  Note that the Froude number is meaningful only for stably stratified flows.  Also, small $\mathrm{Fr}$ implies strongly stratified flow, while strong $\mathrm{Ri}$ indicates strong buoyancy.  

Note that in later discussion we will focus our discussions on Eqs.~(\ref{eq:u_ndim}-\ref{eq:inc_ndim}).  For convenience, we drop the primes from the variables in the subsequent discussions.
 
In some of the earlier studies on {\em strongly} stratified flows, e.g.  Lindborg~\cite{ Lindborg:GRL2005,Lindborg:JFM2006}, Brethouwer et al.~\cite{Brethouwer:JFM2007}, Bartello and Tobias~\cite{Bartello:JFM2013}, the equations  have been written for horizontal and vertical components of the velocity field in terms of  the Froude number and Reynolds number (see Appendix A).   However, we use Eqs.~(\ref{eq:u_ndim}-\ref{eq:inc_ndim}) for our analysis since they help us contrast stably stratified flows and RBC in a single formalism.  In the following discussion we contrast our assumptions and equations with those used for {\em strongly} stratified flows (see Appendix~\ref{sec:appendix}):
\begin{enumerate}[(a)]
\item  A large number of earlier work, e.g. Lindborg~\cite{Lindborg:JFM2006, Lindborg:GRL2005}, Brethouwer et al.~\cite{Brethouwer:JFM2007}, Bartello and Tobias~\cite{Bartello:JFM2013} focus on strongly stratified flows.  A signature of such flows is that their Froude number is much less than unity.  Our focus is on moderately stratified flows, which is achieved by setting the Froude number to unity or higher, or  $u'_\mathrm{rms} \ge 1$ (see Eq.~(\ref{eq:Fr})).  However, $\mathrm{Ri} \le 1$ for such flows. In Sec.~\ref{subsec:stratified} we will show that for $\mathrm{Ri} = O(1)$, a buoyancy dominated flow, we obtain the {\em BO} scaling.  However  for $\mathrm{Ri} \ll 1$, a weakly buoyant flow, we obtain the {\em KO} scaling since the nonlinearity is weak for this case.

\item The strongly stratified flows ($\mathrm{Fr} \ll 1$) are  quasi two-dimensional and strongly anisotropic, hence they employ $L_z \ll L_x, L_y$ (here $L_x,L_y,L_z$ are the lengths of the box along $x,y,z$ directions respectively)~\cite{Lindborg:GRL2005,Lindborg:JFM2006,Brethouwer:JFM2007,Bartello:JFM2013}.  These flows are expected to model the atmosphere of the Earth.  Our flows, however, are three-dimensional and weakly-anisotropic since $\mathrm{Fr} \ge 1$.  Therefore, we simulate the flows in geometries where $L_x \approx  L_y \approx L_z$.  The latter configurations are suitable for testing Bolgiano-Obukhov scaling, which is formulated as an isotropic spectrum.

\item  For the non-dimensionalized Eqs.~(\ref{eq:u_ndim}-\ref{eq:inc_ndim}), the Brunt-V\"ais\"al\"a frequency $N$ is unity, implying that the time scale of the gravity waves is of the same order as the eddy turnover time of the large eddies.

\item Our flows are turbulent, i.e., $\mathrm{Re} \gg 1$.

\item A large number of stably stratified flow simulations (e.g.,  Lindborg~\cite{Lindborg:GRL2005,Lindborg:JFM2006}, Brethouwer et al.~\cite{Brethouwer:JFM2007}, Vallgren et al.~\cite{Vallgren:PRL2011},  Kimura and Herring~\cite{Kimura:JFM2012}, and  Bartello and Tobias~\cite{Bartello:JFM2013}) employ periodic boundary condition; this is to simulate the bulk flow away from the boundaries.  Also, the Bolgiano and Obukhov scaling, as well as Kolmogorov phenomenology, are strictly applicable for homogeneous and isotropic turbulence, for which a periodic box is a good geometrical configuration.  Keeping these aspects in mind, we employ the periodic boundary condition for simulating stably stratified flows. 

Boundary walls and thermal plates play an important role in the flow dynamics of RBC. In our present study, at the top and bottom plates, we employ the free-slip boundary condition for the velocity field, and the conducting boundary condition for the temperature field.  We apply the periodic boundary condition at the side walls.

\end{enumerate}

We simulate the stably stratified flow and RBC by solving Eqs.~(\ref{eq:u_ndim}-\ref{eq:inc_ndim}) numerically for the aforementioned boundary conditions.  After that we study kinetic energy spectrum and flux, as well as other diagnostics tools like energy supply rate by buoyancy; we will discuss these tools in the next section.

\subsection{Energy flux and other diagnostics}

In Fourier space, the equation for the kinetic energy is derived using Eq.~(\ref{eq:u_ndim}) as~\cite{Lesieur:book,Lvov:PRL1991,Verma:EPL2012}
\begin{equation}
\frac{\partial E_u(k)}{\partial t} = T_u(k) + F(k) - D(k),
\label{eq:dEk_dt}
\end{equation}
where $E_u(k)$ is the kinetic energy of the wavenumber shell of radius $k$,  $T_u(k)$ is energy transfer rate to the  shell $k$ due to nonlinear interactions, and $F(k)$ is total energy supply rate to the shell from the forcing functions, both buoyancy and external forcing $\bf f^u$:
 \begin{equation}
F(k) = \sum_{|{\mathbf k}| = k} \Re\langle u_z({\mathbf k}) \theta^*({\mathbf k}) \rangle + \sum_{|{\mathbf k}| = k}  \Re\langle {\mathbf u}({\mathbf k}) \cdot {\mathbf f}^*({\mathbf k}) \rangle,
\label{eq:F} 
\end{equation}
where the first term is due to buoyancy, while the second term is due to the external random forcing.  The term $D(k)$ of Eq.~(\ref{eq:dEk_dt}) is the viscous dissipation rate, and is given by
 \begin{equation}
D(k) = \sum_{|{\mathbf k}| = k} 2 \sqrt{\frac{\mathrm{Pr}}{\mathrm{Ra}}} k^2 E_u(k),
\label{eq:D_k}
\end{equation}
which is always positive.

The nonlinear interaction term $T_u(k)$  is related to the kinetic energy flux $\Pi_u(k)$ as
\begin{equation}
\Pi_u(k) = - \int_0^k T_u(k)\,dk, 
\label{eq:Pi_def}
\end{equation}
which is computed using the following formula~\cite{Verma:PR2004}
\begin{equation}
\Pi_u(k_0)  =  \sum_{k \ge k_0} \sum_{p<k_0} \delta_{\bf k,\bf p+ \bf q} \Im([{\bf k \cdot u(q)}]  [{\bf u^*(k) \cdot u(p)}])
\label{eq:Pi_Verma}							
\end{equation}
The energy flux $\Pi_u(k_0) $ is interpreted as the kinetic energy leaving a wavenumber sphere of radius $k_0$.

\begin{figure}[htbp]
\begin{center}
\includegraphics[scale = 1]{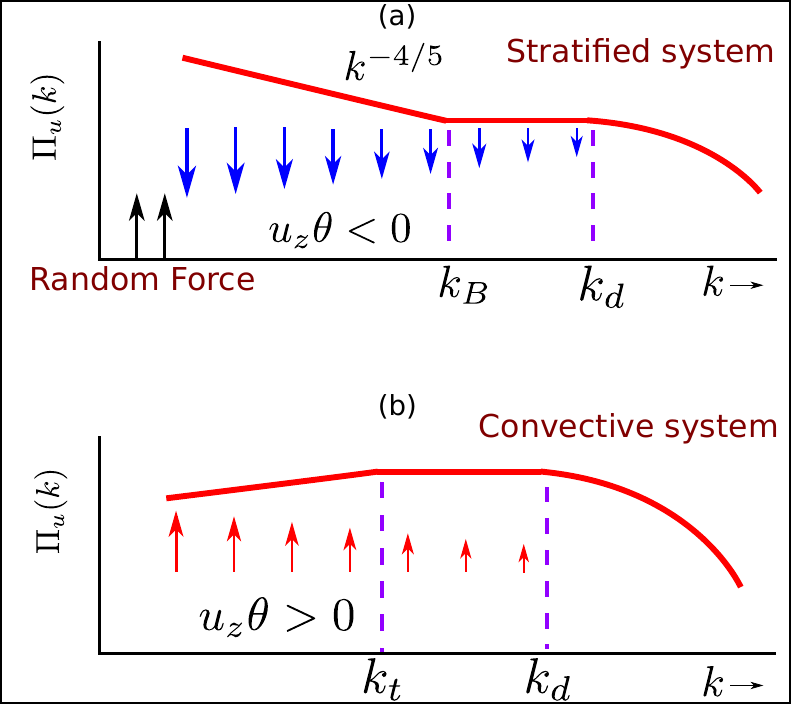}
\end{center}
\caption{Schematic diagrams of energy flux $\Pi_u(k)$: (a) In a stably stratified flow,  $\Pi_u(k)$ decreases with $k$  due to a negative energy supply rate $\Re \langle u_z(k) \theta^*(k)\rangle$; (b) In Rayleigh B\'{e}nard convection, $\Re \langle u_z(k) \theta^*(k) \rangle >0$, hence  $\Pi_u(k)$ first increases for  $k < k_t$ where $F(k) > D(k)$, then $\Pi_u(k) \approx \mathrm{const}$ $k_t < k < k_d$ where $F(k) \approx D(k)$; $\Pi_u(k)$ decreases for $k>k_d$ where $F(k) < D(k)$. }
\label{fig:sch_flux}
\end{figure}

Using Eqs.~(\ref{eq:dEk_dt},\ref{eq:Pi_def}), we deduce that 
\begin{equation}
\frac{d}{dk}\Pi_u(k) = -T_u(k) = -\frac{\partial E_u(k)}{\partial t} + F(k) - D(k).
\label{eq:dPi}
\end{equation}
Under a steady state ($\partial E_u(k)/\partial t =0$), we obtain 
\begin{equation}
\frac{d}{dk} \Pi_u(k) =  F(k) - D(k)
\label{eq:dPik_dk}
\end{equation}
or
\begin{equation}
\Pi_u(k+\Delta k) =  \Pi_u(k) + ( F(k) - D(k)) \Delta k.
\label{eq:Pi_k_Fk_Dk}
\end{equation}
Equation (\ref{eq:Pi_k_Fk_Dk}) is obvious, but it provides us important clues on the energy spectrum and flux of the buoyancy-driven flows.  Here we list three possibilities for the inertial range ($k_f < k < k_d$), where $k_f$ is the forcing wavenumber, and $k_d$ is the dissipation wavenumber:
\begin{enumerate}

\item  For the inertial range of fluid turbulence, $F(k)=0$  and $D(k) \rightarrow 0$, hence $\Pi_u(k+\Delta k) \approx  \Pi_u(k)$ and $E_u(k) \sim k^{-5/3}$, which is the prediction of Kolmogorov's theory.
\label{enu_1}

\item For the stably stratified flows ($S=-1$ in Eq.~(\ref{eq:th_ndim})),  as argued by Bolgiano and Obukhov, the buoyancy converts kinetic energy of the flow to potential energy, i.e., $F(k) =\Re \langle u_z(k) \theta^*(k) \rangle < 0$ for $k_f < k< k_B$.  Therefore, Eq.~(\ref{eq:Pi_k_Fk_Dk}) predicts that $\Pi_u(k)$ will decrease with $k$ in this wavenumber range, as shown in Fig.~\ref{fig:sch_flux}{\color{blue}(a)}. In the wavenumber range, $k_B < k < k_d$, buoyancy becomes weaker, hence $\Pi_u(k) \sim \mathrm{const}$, and Kolmogorov's spectrum is expected.  In the present paper, using numerical simulation, we demonstrate {\em BO} scaling  in the $k_f < k< k_B$ regime; the demonstration of  $E_u(k) \sim k^{-5/3}$ for $k_B < k < k_d$ requires larger resolution than that used in this paper.
\label{enu_2}

\item For RBC ($S=1$ in Eq.~(\ref{eq:th_ndim})), buoyancy feeds energy to the kinetic energy, hence  $F(k) =  \Re \langle u_z(k) \theta^*(k) \rangle > 0$.  Therefore, the sign of $d\Pi_u(k)/dk$ depends crucially on $D(k)$.  First, for  $k < k_t$, $\Pi_u(k)/dk > 0$ since $F(k) > D(k)$, then for the intermediate wavenumbers $k_t < k < k_d$ where $F(k) \approx D(k)$, we expect $\Pi_u(k)/dk \approx 0$.  Finally in the dissipative range ($k>k_d$), $\Pi_u(k)/dk < 0$ since $F(k) < D(k)$. Here  $k_t$ is the transition wavenumber shown in Fig.~\ref{fig:sch_flux}{\color{blue}(b)}.  Consequently, as shown in Fig.~\ref{fig:sch_flux}{\color{blue}(b)}, the flux $\Pi_u(k)$ first increases, then flattens, and finally decreases, in the three wavenumber bands discussed above.  In the intermediate band, $k_t < k < k_d$, we observe Kolmogorov's $k^{-5/3}$ spectrum due to a constant KE flux.

 Since the flux does not decrease due to buoyancy (see Eq.~(\ref{eq:pi})),  the {\em BO} scaling is not applicable to RBC turbulence,  contrary to the predictions by Procaccia and Zeitak~\cite{Procaccia:PRL1989}, L'vov~\cite{Lvov:PRL1991},  L'vov and Falkovich~\cite{Lvov:PD1992}, and Rubinstein~\cite{Rubinstein:NASA1994}.

\label{enu_3}
\end{enumerate}

There is another useful flux called the entropy flux $\Pi_\theta$, which is defined as
\begin{equation}
\Pi_\theta(k_0)  =  \sum_{k \ge k_0} \sum_{p<k_0}  \delta_{\bf k,p+q} \Im([{\bf k \cdot u(q)}]  [{\bf \theta^*(k) \cdot \theta(p)}]).
\end{equation}
Both, the {\em KO} and {\em BO}, phenomenologies predict a constant $\Pi_\theta$.

In this paper we simulate stably stratified  flows and RBC, and compute  the kinetic energy and entropy spectra, as well as fluxes.  We also compute $F(k), D(k)$, and $d\Pi_u(k)/dk$, and show that our results are in good agreement with the arguments of  items~\ref{enu_2} and~\ref{enu_3} discussed above.  For stably stratified flows, the {\em BO} scaling is observed when $\mathrm{Ri} = O(1)$, but the Kolmogorov scaling $E_u(k) \sim k^{-5/3}$ is observed when $\mathrm{Ri} \ll 1$, or when buoyancy is negligible.   RBC flows, however, exhibit  the Kolmogorov scaling $E_u(k) \sim k^{-5/3}$ for a narrow band of wavenumbers.

\section{Simulation Method}
\label{sec:method}
We perform direct numerical simulation of stably stratified flows and RBC in a three-dimensional box by solving Eqs.~(\ref{eq:u_ndim}-\ref{eq:inc_ndim}) using pseudospectral code Tarang~\cite{Verma:Pramana2013}.    We employ fourth-order Runge-Kutta (RK4) method  for time stepping,  Courant-Freidricks-Lewey (CFL) condition for computing time step $\Delta t$, and $3/2$ rule  for dealiasing.  

For the stratified flows, we employ the periodic boundary conditions on all sides of a cubic box of size $(2\pi)^3$.   To obtain a steady turbulent flow, we apply a random force to the flow in the band $2 \le k \le 4$ using the scheme of Kimura and Herring~\cite{Kimura:JFM2012}.  The parameters chosen for our simulations are  $\mathrm{Pr} =1.0$ (close to that of air), and  Richardson numbers  $\mathrm{Ri}=4 \times 10^{-7}, 0.01$, and  $0.5$.  The grid resolution for $\mathrm{Ri}= 0.01$ is  $1024^3$, which is one of the largest grids for such simulations.  The resolutions for $\mathrm{Ri}= 4 \times 10^{-7}$ and $0.5$ are $512^3$ grids.  The parameters of our runs are listed in Table~\ref{table:simulation_details}. All our simulations are fully resolved since $k_{\mathrm max} \eta >1$, where $k_{\mathrm max}$ is the maximum wavenumber of the run, and $\eta$ is the Kolmogorov length scale.

We simulate RBC of a fluid in a unit box with $512^3$ grid.  The parameters of the simulation are $\mathrm{Pr} = 1$ and Rayleigh number $\mathrm{Ra} =10^7$.   For the horizontal plates, we employ free-slip boundary condition for the velocity field, and conducting boundary condition, i.e. $\theta=0$, for the temperature field.  For the vertical walls, we apply periodic boundary condition for both the fields.   Simulation details of RBC simulation  are listed at the bottom row of the Table~\ref{table:simulation_details}. 
 
In the next section we will compute the the spectra and fluxes of the kinetic energy as well as that of entropy.

\begin{table*}[htbp]
\begin{center}
\caption{Parameters of our numerical simulations for stably stratified flow (first three rows), and Rayleigh B\'{e}nard convection (the last row): Grid size, Richardson number $\mathrm{Ri}$, Rayleigh number $\mathrm{Ra}$, Reynolds number $\mathrm{Re}$, Froude
number $\mathrm{Fr}$, kinetic energy dissipation rate $\epsilon_u$, entropy dissipation rate $\epsilon_{\theta}$, Anisotropy ratio ${E_\perp}/{2E_\parallel}$, where $E_\perp = (u_x^2 + u_y^2)/2$ and $E_\parallel = u_z^2/2$, $k_{max}\eta$ where $\eta$ is the Kolmogorov length, Bolgiano wavenumber $k_B$, and averaged $\Delta t$.  We choose $\mathrm{Pr}=1$ for all the runs.}
\begin{tabular}{p{1.0cm} p{1.4cm} p{1.2cm} p{1.2cm} p{1.6cm} p{1.6cm} p{1.6cm} p{1.2cm} p{1.0cm} p{1.0cm} p{1.6cm}}
\hline \hline \\[0.3 pt]
Grid & $\mathrm{Ri}$ & $\mathrm{Ra}$ & $\mathrm{Re}$ & $\mathrm{Fr}$  &  $\epsilon_u$ & $\epsilon_{\theta}$ & ${E_\perp}/{2E_\parallel}$ & $k_{max} \eta$ & $k_B$ & $\Delta t$\\[2 mm]
\hline \\[0.5 pt]
$512^3$ & $0.5$ & $1 \times 10^5$ & $467$ &$1.4$& $0.47$ & $60.7$ & $1.2$ &$4.2$  & $6.0$ & $2.5 \times 10^{-5}$\\
$1024^3$ & $0.01$ & $5 \times 10^3$  &$649$ & $10$ & $114$ & $150$ & $1.0$ &$6.4$  & $8.5$ & $3.5 \times 10^{-6}$\\
$512^3$ & $4 \times 10^{-7}$ & $0.1$ &$510$ & $1.5 \times 10^3$ & $6.7 \times 10^8$ & $141$ & $1.0$  &$3.8$ & $< 1$ & $2.6 \times 10^{-6}$\\
\hline \\
$512^3$ & $16$ & $10^7$ & $790$ & NA & $8.8 \times 10^{-3}$ & $1.0 \times 10^{-3}$ & $0.41$ &$2.6$ & NA &  $6.2 \times 10^{-4}$\\
\hline \hline
\end{tabular}
\label{table:simulation_details}
\end{center}
\end{table*}

\section{Numerical results}
\label{sec:results}

We compute the the spectra and fluxes of the kinetic energy as well as that of entropy using the steady-state data.  We will also compute $F(k), D(k)$, and $d\Pi_u(k)/dk$ for the flows.  These results will be discussed below.

\subsection{Stably Stratified Flow}
\label{subsec:stratified}
\begin{figure}[htbp]
\begin{center}
\includegraphics[scale = 1.0]{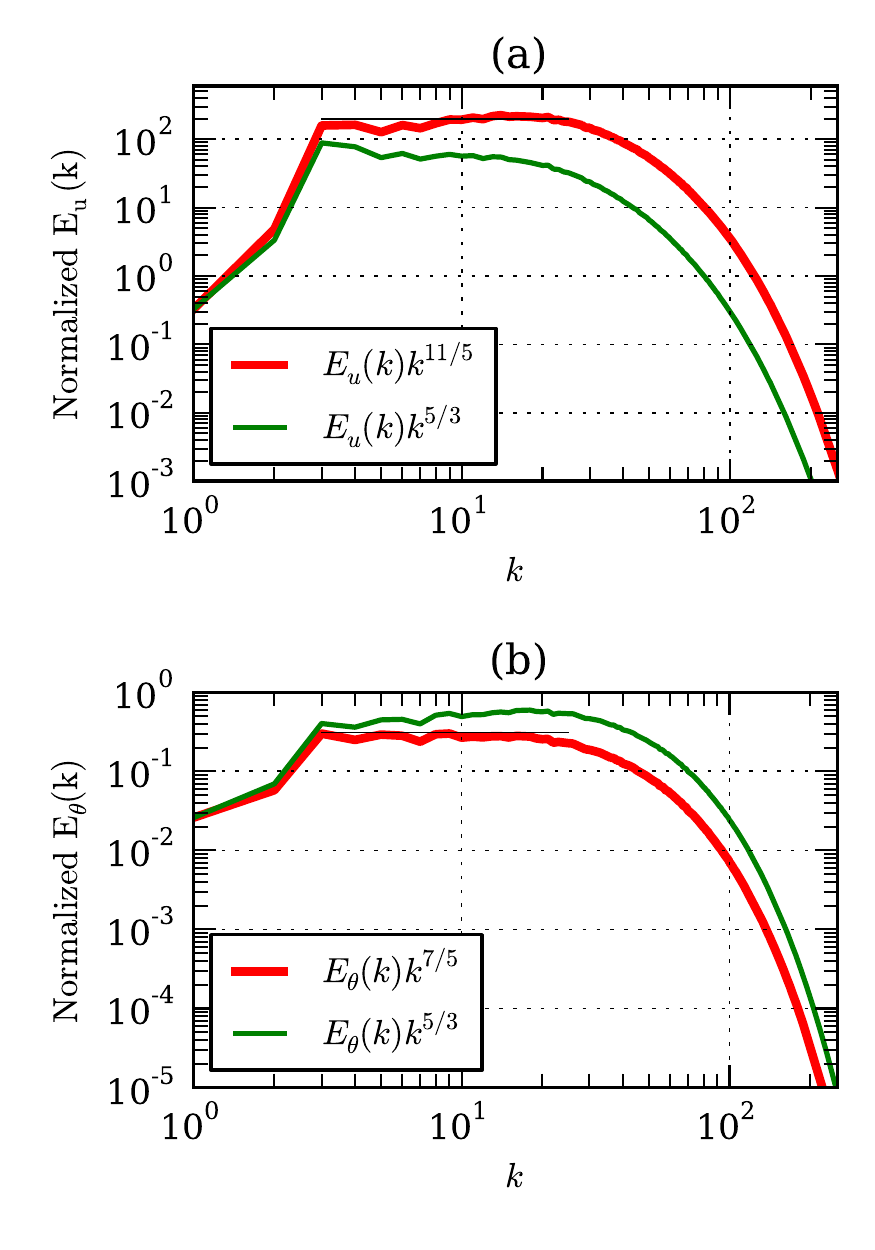}
\end{center}
\setlength{\abovecaptionskip}{0pt}
\caption{For stably stratified simulation with $\mathrm{Pr}=1$ and $\mathrm{Ri} = 0.01$, plots of (a) normalized KE and (b) entropy spectra for Bolgiano-Obukhov ({\em BO}) and Kolmogorov-Obukhov ({\em KO}) scaling.  {\em BO} scaling fits with the data better than {\em KO} scaling.}
\label{fig:spectra_0_01}
\end{figure}

First, we simulate stably stratified flows for $\mathrm{Pr}=1$ and $\mathrm{Ri}=0.01$ on a $1024^3$ grid, and compute the spectrum and flux using the steady state data.   Fig.~\ref{fig:spectra_0_01}{\color{blue}(a)} illustrates the normalized KE spectra, $E_u(k)k^{11/5}$ for the {\em BO} scaling, and $E_u(k)k^{5/3}$ for the {\em KO} scaling.  The numerical data fits with the {\em BO} scaling quite well for approximately a decade, thus confirming the phenomenology of Bolgiano and Obukhov.  The normalized entropy spectra, $E_\theta(k)k^{7/5}$ ({\em BO} scaling) and $E_\theta(k)k^{5/3}$ ({\em KO} scaling), illustrated in Fig.~\ref{fig:spectra_0_01}{\color{blue}(b)} also show that the {\em BO} scaling is preferred for $\mathrm{Ri}=0.01$ stably stratified flow. 

\begin{figure}[htbp]
\begin{center}
\includegraphics[scale = 1.0]{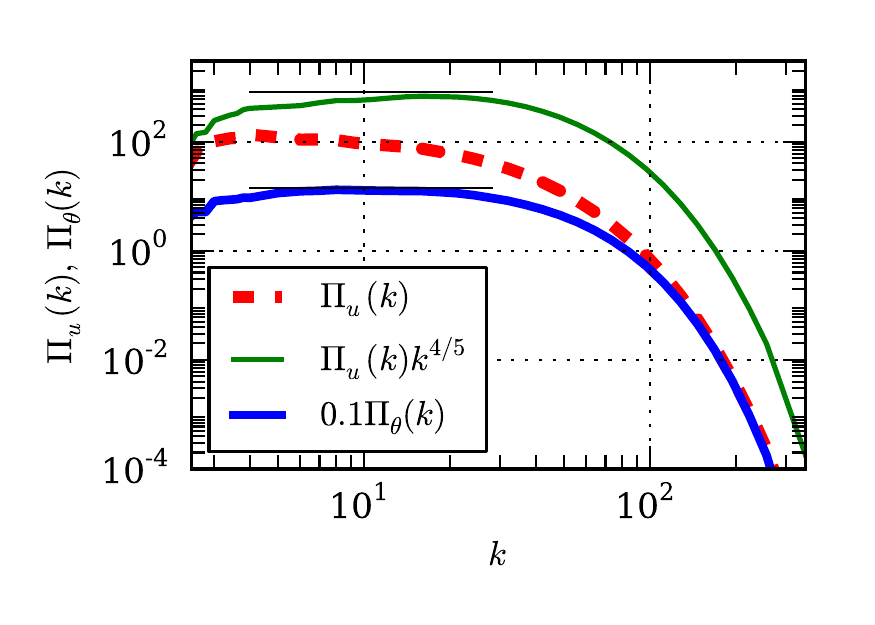}
\end{center}
\setlength{\abovecaptionskip}{0pt}
\caption{For stably stratified simulation with $\mathrm{Pr}=1$ and $\mathrm{Ri} = 0.01$, plots of KE flux $\Pi_u(k)$, normalized KE flux $\Pi_u(k)k^{4/5}$, and entropy flux $\Pi_{\theta}(k)$.}
\label{fig:flux}
\end{figure}

We cross check our spectrum results with the KE and entropy fluxes, which are plotted in Fig.~\ref{fig:flux}.   Clearly, the KE flux, $\Pi_u(k)$, is positive, and it decreases with $k$.  However  $\Pi_u(k) k^{4/5}$ is almost flat, thus  $\Pi_u(k) \propto k^{-4/5}$, same as Eq.~(\ref{eq:pi}).  We also observe that $\Pi_\theta$ is a constant in the inertial range [Eq.~(\ref{eq:pi_theta})]; thus flux results are consistent with the {\em BO} predictions.  

We also compute the Bolgiano wavenumber $k_B$~\cite{Bolgiano:JGR1959} using the numerical data, and find that $k_B \approx 8.5$.  Our plots on spectra and fluxes show that  $k_B \approx 8.5$ is only 3 to 4 times smaller than $k_d$, wavenumber where the dissipation range starts.  Therefore a clear-cut crossover from $k^{-11/5}$ to $k^{-5/3}$ is not observed in our simulations.   We are in the process of performing simulations on even higher resolution to probe the dual spectra ($k^{-11/5}$ and $k^{-5/3}$). 

\begin{figure}[htbp]
\begin{center}
\includegraphics[scale = 1.0]{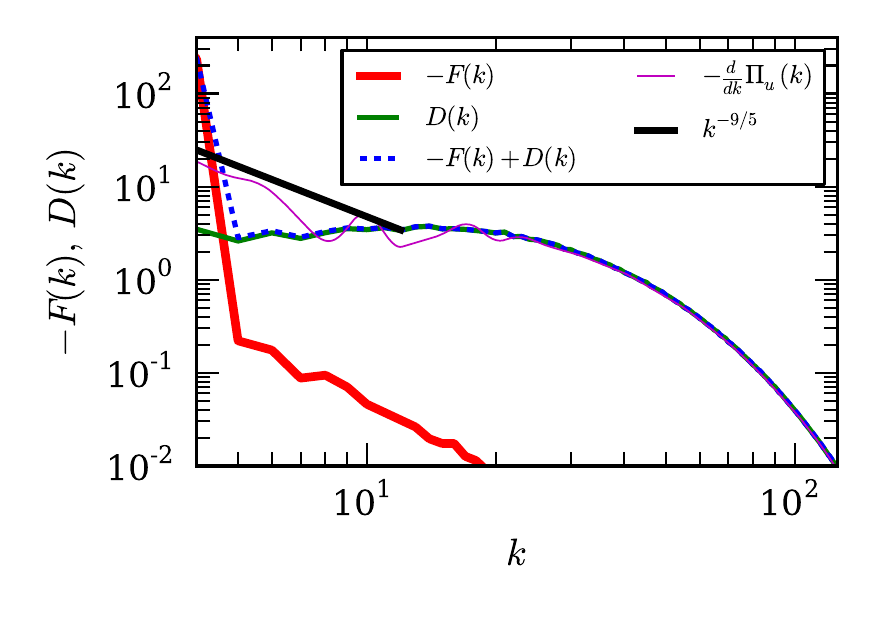}
\end{center}
\setlength{\abovecaptionskip}{0pt}
\caption{For stably stratified simulation with $\mathrm{Pr}=1$ and $\mathrm{Ri}=0.01$, plots of $-F(k) , D(k), [-F(k)+D(k)]$, $-d\Pi_u(k)/dk$, and $k^{-9/5}$ line to match with  $-d\Pi_u(k)/dk$ in the small $k$ regime.}
\label{fig:deriv_1}
\end{figure}

We also compute energy supply rate by buoyancy, $F(k) =  \Re \langle u_z(k) \theta^*(k)\rangle$, $D(k)$, and $d\Pi_u(k)/dk$ using the numerical data, and plot them in Fig.~\ref{fig:deriv_1}.  The figure illustrate that $F(k) < 0$, as argued in item~\ref{enu_2} of Sec.~\ref{sec:gov_eqns}.  The negative $F(k)$ implies that $\Pi_u(k)$ decreases with $k$ even without $D(k)$, which is a crucial ingredient for the {\em BO} scaling.   Note that the kinetic energy flux is depleted by both $F(k)$ and $D(k)$, and they satisfy the relation of Eq.~(\ref{eq:dPik_dk}).  Interestingly, for small $k$, $d\Pi_u(k)/dk \sim k^{-9/5}$ (the black line of Fig.~\ref{fig:deriv_1}), consistent with $\Pi_u(k) \sim k^{-4/5}$.

\begin{figure}[htbp]
\begin{center}
\includegraphics[scale = 0.9]{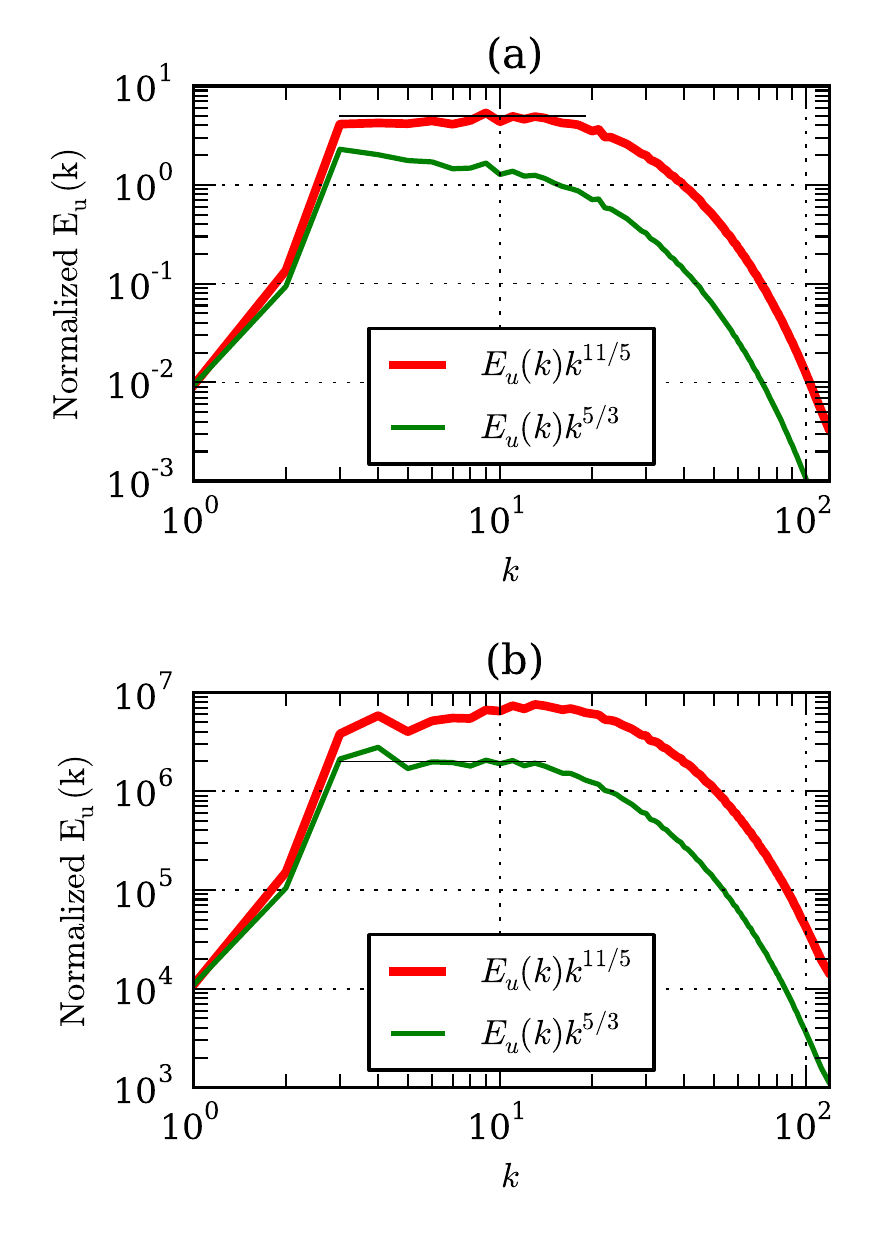}
\end{center}
\setlength{\abovecaptionskip}{0pt}
\caption{For stably stratified simulation with $\mathrm{Pr}=1$, and (a) $\mathrm{Ri} = 0.5$ and (b) $\mathrm{Ri} = 4 \times 10^{-7}$, the plots of normalized KE spectra for Bolgiano-Obukhov ({\em BO}) scaling and Kolmogorov-Obukhov ({\em KO}) scaling. }
\label{fig:ke_spec}
\end{figure}

We also performed $512^3$ grid simulations for ${\rm Ri}= 0.5 $ and $4 \times 10^{-7}$ with $\mathrm{Pr}=1$.  The normalized KE spectra for these two cases are exhibited in Figs.~\ref{fig:ke_spec}{\color{blue}(a)} and ~\ref{fig:ke_spec}{\color{blue}(b)} respectively.  Our results show that {\em BO} scaling is valid for ${\rm Ri} = 0.5$, but {\em KO} scaling (with a constant $\Pi_u(k)$) is valid for ${\rm Ri } = 4 \times 10^{-7} $, which is as expected since buoyancy is significant only for moderate and large  ${\rm Ri}$'s.

\begin{figure}[htbp]
\begin{center}
\includegraphics[scale = 0.9]{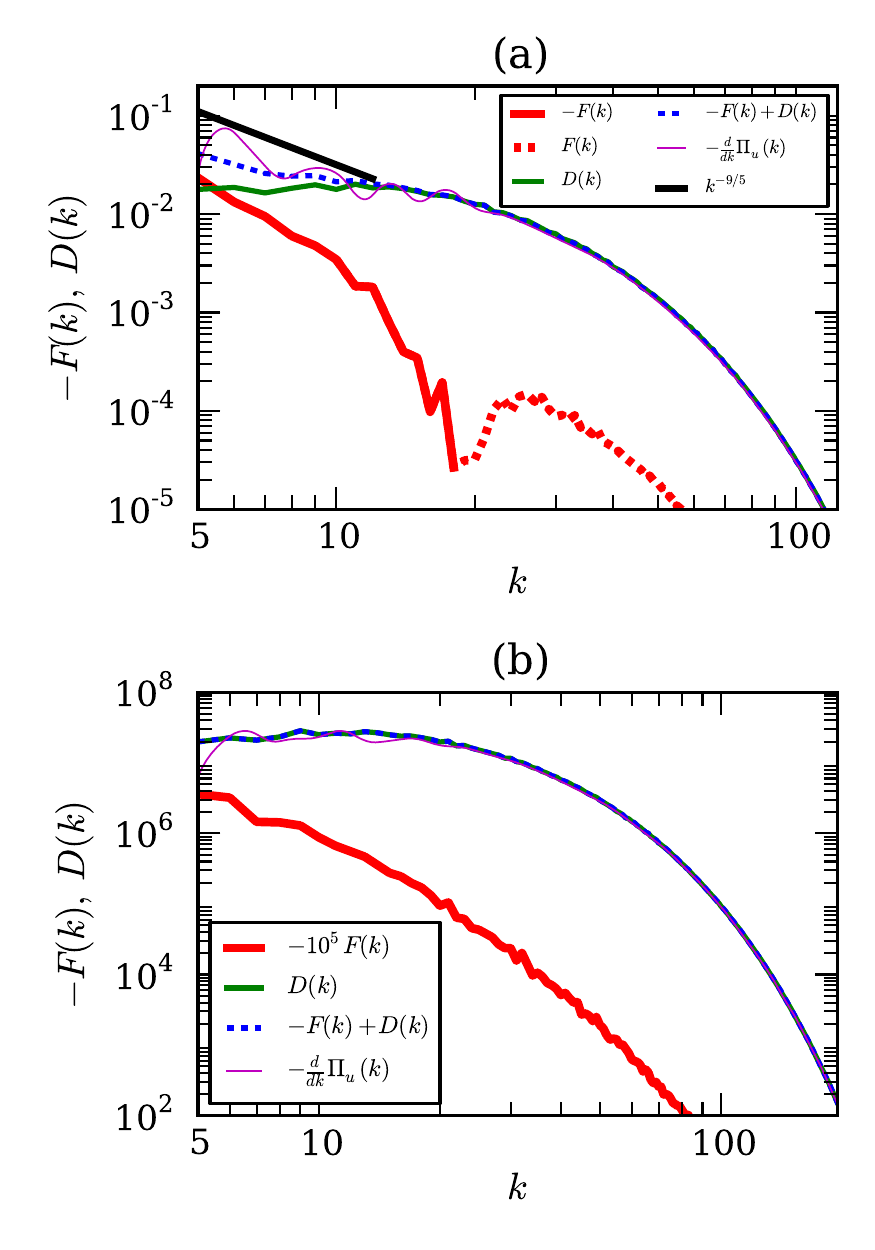}
\end{center}
\setlength{\abovecaptionskip}{0pt}
\caption{For stably stratified simulation with $\mathrm{Pr}=1$, and (a) $\mathrm{Ri} = 0.5$ and (b) $\mathrm{Ri} = 4 \times 10^{-7}$,   plots of $-F(k) , D(k), [-F(k)+D(k)]$, $-d\Pi_u(k)/dk$, and $k^{-9/5}$ line to match with  $-d\Pi_u(k)/dk$ in the small $k$ regime.  In Fig.~(a), the negative $F(k)$ is shown as solid red curve, and positive $F(k)$ as dashed red curve. In Fig.~(b), $F(k)$ is multiplied by $10^5$ to fit in the same range.}
\label{fig:deriv_2}
\end{figure}

We compute $F(k)$, $D(k)$, and $d\Pi_u(k)/dk$ for $\mathrm{Ri}= 0.5$ and $4\times 10^{-7}$, and plot them in Figs.~\ref{fig:deriv_2}{\color{blue}(a,b)} respectively.  In the inertial range, $F(k) < 0 $ for both the cases, just like $\mathrm{Ri}= 0.01$.  The behaviour of $F(k)$, $D(k)$, and $d\Pi_u(k)/dk$ for $\mathrm{Ri}= 0.5$ is very similar to that of $\mathrm{Ri}= 0.01$, except that $F(k)$ for $\mathrm{Ri}= 0.5$ is a bit smaller than that for $\mathrm{Ri}= 0.01$.   For $\mathrm{Ri} = 4\times 10^{-7}$, buoyancy is weak, hence $F(k)$  is much smaller than that for $\mathrm{Ri}= 0.01$, which leads to an approximately constant $\Pi_u(k)$, and Kolmogorov's spectrum for the kinetic energy.

\begin{figure}[htbp]
\begin{center}
\includegraphics[scale = 0.9]{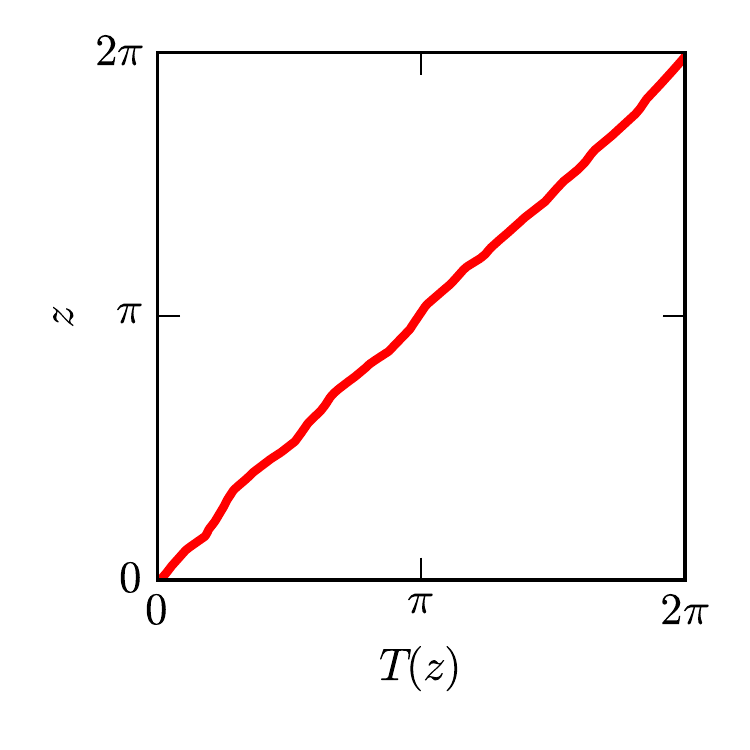}
\end{center}
\setlength{\abovecaptionskip}{0pt}
\caption{For stably stratified simulation with $\mathrm{Pr}=1$ and $\mathrm{Ri}=0.01$, the vertical variation of horizontally averaged mean temperature $\bar{T}(z) = \langle T(x,y,z) \rangle_{xy}$.}
\label{fig:temp_pro}
\end{figure}

Recall that we employ periodic boundary condition for the stably stratified flows in the vertical direction, thus eliminating the effects of  boundary walls.  In Fig.~\ref{fig:temp_pro} we plot the plane-averaged (over $xy$ plane) mean temperature profile $\bar{T}(z) = \langle T(x,y,z) \rangle_{xy}$.   Since $\bar{T}(z) $ is linear,  a constant temperature gradient $d\bar{T}/dz$ (hence buoyancy) acts in the whole box.  Therefore,  {\em BO} scaling is expected everywhere.   It is important to contrast the above profile with that for Rayleigh-B\'{e}nard convection in which most of the temperature drop takes place in the narrow thermal boundary layers at the plates~\cite{Moore:JFM1973,Verzicco:JFM2003}, while the bulk flow has $d\bar{T}/dz \approx 0$.  Thus we expect {\em BO} scaling in the boundary layers, and {\em KO} scaling in the bulk, as reported by Calzavarini et al.~\cite{Calzavarini:PRE2002}.

In the next subsection we will discuss the results of Rayleigh B\'{e}nard Convection.

\subsection{Rayleigh B\'{e}nard Convection}

Borue and Orszag~\cite{Borue:JSC1997}, and \u{S}kandera et al.~\cite{Skandera:HPCISEG2SBH2009}  simulated RBC flow under periodic boundary condition. They observed the {\em KO} scaling for both velocity and temperature fields, consistent with the arguments presented in Sec.~\ref{sec:gov_eqns}.  A shell model approximates the turbulence in a periodic box quite well; a recent shell model of RBC flow~\cite{Kumar:arxiv2014_2} also yields  {\em KO} scaling, consistent with the numerical results of Borue and Orszag~\cite{Borue:JSC1997}, and \u{S}kandera et al.~\cite{Skandera:HPCISEG2SBH2009}.   In a typical RBC flow, however, a fluid is confined between two horizontal conducting plates that are maintained at  constant temperatures, with the bottom plate hotter than the top one.   Earlier, Mishra and Verma~\cite{Mishra:PRE2010} showed that zero- and small Prandtl number RBC exhibit Kolmgorov's spectrum for the kinetic energy, but their results were inconclusive for moderate Prandtl number RBC.  In this subsection, we will investigate this issue for $\mathrm{Pr}=1$.

\begin{figure}[htbp]
\begin{center}
\includegraphics[scale = 0.9]{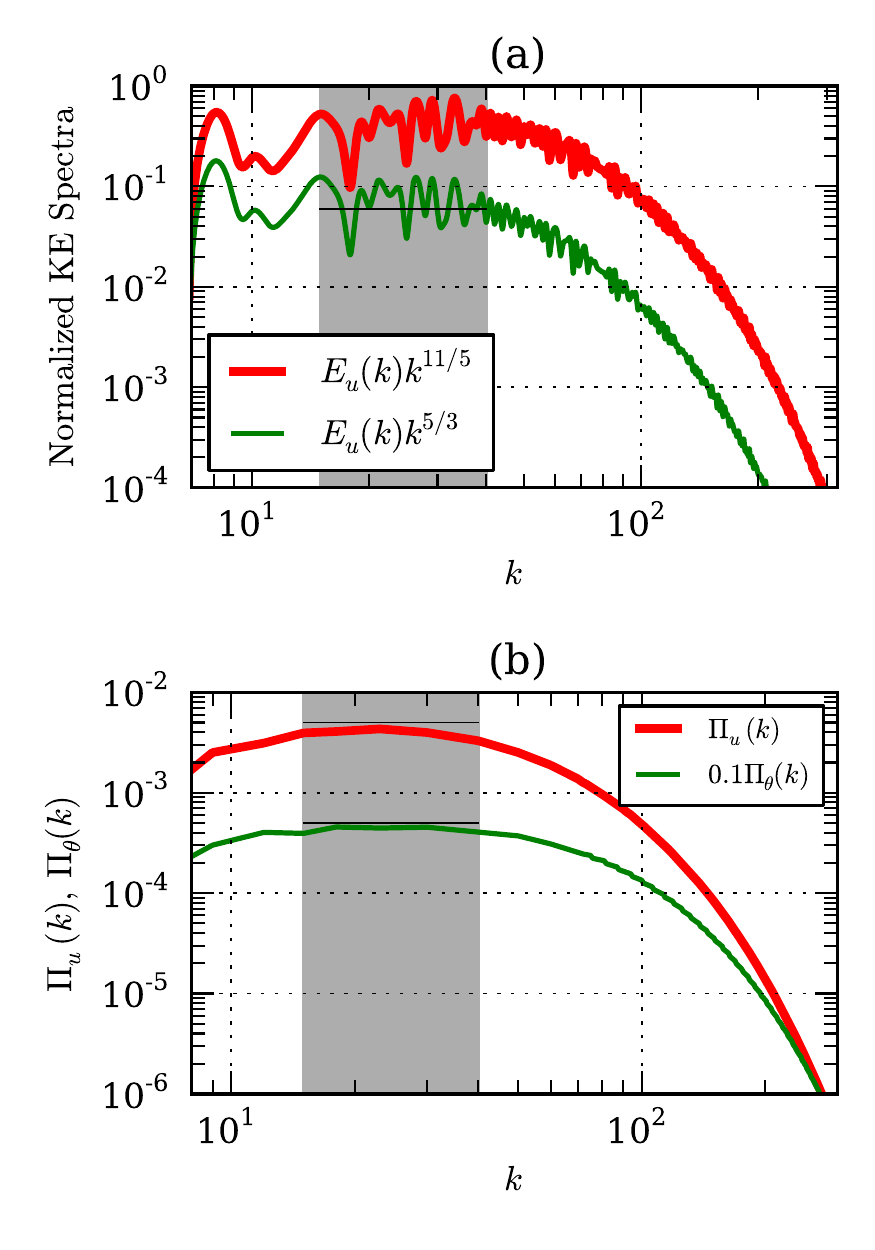}
\end{center}
\setlength{\abovecaptionskip}{0pt}
\caption{For RBC simulation with $\mathrm{Pr}=1$ and $\mathrm{Ra}=10^7$, (a) plots of normalized KE spectra for Bolgiano-Obukhov ({\em BO}) and Kolmogorov-Obukhov ({\em KO}) scaling; {\em KO} scaling fits with the data better than {\em BO} scaling;(b) KE flux $\Pi_u(k)$ and entropy flux $\Pi_\theta(k)$. The shaded region shows the inertial range.}
\label{fig:ke_specra_rbc}
\end{figure}

 To explore which of the two scaling ({\em KO} or {\em BO}) is applicable for  RBC turbulence with plates, we perform RBC simulations   for  $\mathrm{Pr}=1$ and $\mathrm{Ra} = 10^7$,  and  compute the spectra and fluxes of the KE as well as the entropy for the steady state data.  In Fig.~\ref{fig:ke_specra_rbc}{\color{blue}(a)}, we plot the normalized KE spectra for the {\em BO} and the {\em KO} scaling.  The plots indicate that the {\em KO} scaling fits better than the {\em BO} scaling for a narrow band of wavenumbers (the shaded region, $15 < k < 40$).  
 
\begin{figure}[htbp]
\begin{center}
\includegraphics[scale = 0.9]{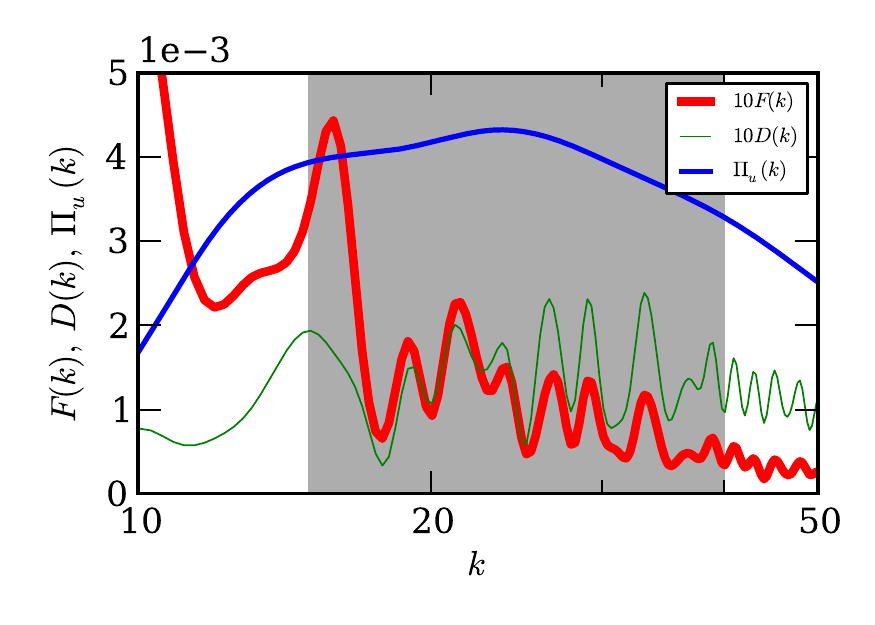}
\end{center}
\setlength{\abovecaptionskip}{0pt}
\caption{For RBC simulation with $\mathrm{Pr}=1$ and $\mathrm{Ra}=10^7$, plots of $\Pi_u(k)$, $F(k)$, and $D(k)$ for $10 \le k \le 50$.}
\label{fig:flux_details}
\end{figure}

 We plot the KE and entropy fluxes in Fig~\ref{fig:ke_specra_rbc}{\color{blue}(b)}.   We also plot a zoomed view of the energy flux in Fig.~\ref{fig:flux_details}, according to which KE flux increases till $k=22$, and then starts to decrease.  In the logarithmic scale, the KE flux is an approximate constant for  the wavenumbers $15 < k < 40$, a band where $E_u(k) \sim k^{-5/3}$.  Thus we claim that convective turbulence exhibits Kolmogorov's power law for a narrow band of wavenumbers.   Interestingly, the energy spectrum of RBC  exhibits stronger fluctuations than that of stably stratified turbulence; this feature is possibly due to the ``plumes" emanating from the plates. This feature as well as a larger range of wavenumber exhibiting {\em KO} scaling may be visible in a large resolution simulation, which is planned as a future study.

\begin{figure}[htbp]
\begin{center}
\includegraphics[scale = 0.9]{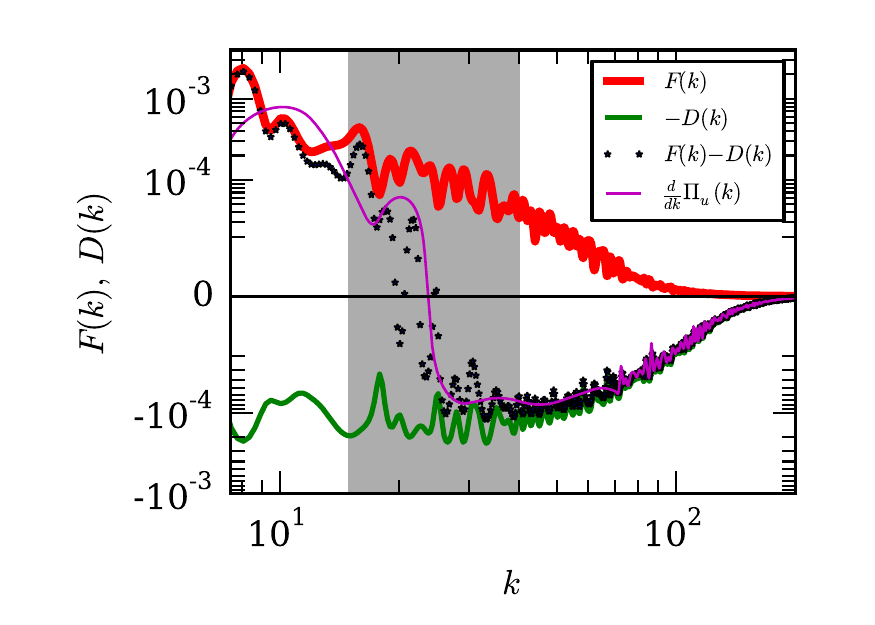}
\end{center}
\setlength{\abovecaptionskip}{0pt}
\caption{For RBC simulation with $\mathrm{Pr}=1$ and $\mathrm{Ra} = 10^7$, plots of $F(k)$, $-D(k)$, $F(k)-D(k)$, and $d\Pi_u(k)/dk$. $d\Pi_u(k)/dk > 0$ for $k < 22$, but $<0$ for $k > 22$. }
\label{fig:deriv_RBC}
\end{figure}

Further investigations of $F(k)$, $\Pi_u(k)$, and $d\Pi_u(k)/dk$ provide stronger evidence for the {\em KO} scaling in RBC.  We plot these quantities in Figs.~\ref{fig:flux_details} and \ref{fig:deriv_RBC}, according to which $F(k) > 0$, consistent with the discussion of Sec.~\ref{sec:gov_eqns} and Fig.~\ref{fig:sch_flux}{\color{blue}(b)}. In addition, for the wavenumber band $7 < k < 22$, $F(k) > D(k)$, hence, according to Eq.~(\ref{eq:dPik_dk}), $d\Pi_u(k)/dk > 0$. Therefore, $\Pi_u(k)$ increases in this band of wavenumbers, as illustrated in Fig.~\ref{fig:flux_details}.  But for $k > 22$, we find that $D(k) > F(k)$  leading to $d\Pi_u(k)/dk < 0$, therefore, $\Pi_u(k)$ decreases with $k$ for this range of $k$.  However, for a narrow band of wavenumbers   $15 < k < 40$, $F(k) \approx D(k)$, hence   $d\Pi_u(k)/dk \approx 0$ or $\Pi_u(k) \approx \mathrm{const}$. The constancy of $\Pi_u(k)$ yields $E_u(k) \sim k^{-5/3}$, consistent with the energy spectrum plots of Fig.~\ref{fig:ke_specra_rbc}.  Note that many simulations, including Mishra and Verma~\cite{Mishra:PRE2010}, reported that $\Pi_u(k) \sim k^{-4/5}$ for moderate $\mathrm{Pr}$, but the decrease of  $\Pi_u(k)$ in their work is essentially due to $D(k)$, not due to buoyancy.

 Thus, the flux and energy supply due to buoyancy reveal  that convective turbulence follows {\em KO} scaling, at least for a narrow range of wavenumbers. The {\em BO} scaling is ruled out for RBC since $F(k) > 0$.

\begin{figure}[htbp]
\begin{center}
\includegraphics[scale = 0.9]{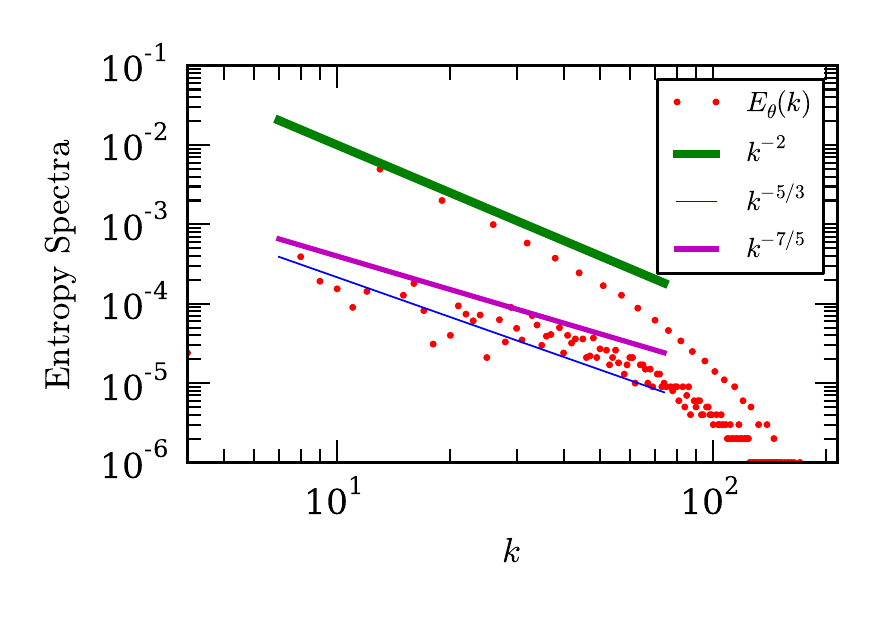}
\end{center}
\setlength{\abovecaptionskip}{0pt}
\caption{For RBC simulation with $\mathrm{Pr}=1$ and $\mathrm{Ra}=10^7$, plots of the entropy spectrum, that exhibits dual branch. The upper branch matches with $k^{-2}$ quite well, while the lower part is fluctuating. }
\label{fig:pe_specra_rbc}
\end{figure}

The entropy $(\theta^2/2)$ is a useful quantity in RBC.  The entropy flux, illustrated in Fig.~\ref{fig:ke_specra_rbc}{\color{blue}(b)},  is constant for the narrow inertial range $(15 < k < 40)$.  In Fig.~\ref{fig:pe_specra_rbc}, we plot the  entropy spectrum that exhibits dual branch, with the upper branch scaling as $k^{-2}$.  Mishra and Verma~\cite{Mishra:PRE2010}, and Pandey et al.~\cite{Pandey:PRE2014} showed the dominant temperature modes $\theta(0,0,2n)$, which are  approximately $ -1/(2 n \pi)$ where $n$ is an integer, constitute the $k^{-2}$ branch of the entropy spectrum. They showed that $\theta(0,0,2n)$ modes are responsible for the steep temperature variations in the thermal boundary layers of the plates.  Interestingly, the temperature modes in both the branches of the entropy spectrum participate to yield a constant entropy flux in the inertial range.

\section{Conclusions}
\label{sec:conclusion}
We performed  large resolution simulations  of stably stratified flows and Rayleigh B\'{e}nard convection, and studied the spectra and fluxes of the kinetic energy and entropy.  We also compute the energy supply rate due to buoyancy that provide important clues on the underlying  turbulence phenomena. 

For stably stratified turbulence, we show that the kinetic energy spectrum $E_u(k) \sim k^{-11/5}$,  the energy flux $\Pi_u(k) \sim k^{-4/5}$, the entropy spectrum $E_\theta(k) \sim k^{-7/5}$, and the entropy flux $\Pi_\theta(k) \sim \mathrm{const}$, in agreement with the prediction of Bolgiano and Obukhov, referred to as {\em BO} scaling.  We also compute the energy supply rate by buoyancy, and find that to be negative, signalling the buoyancy-induced conversion of kinetic energy to  potential energy.  

For the Rayleigh B\'{e}nard convection, the energy supply rate due to buoyancy, $F(k)$, is positive. Hence the kinetic energy flux $\Pi_u(k) $  first increases with $k$, and then flattens for a narrow band of wavenumbers, and finally decreases with $k$; the three regimes correspond to $F(k) > D(k)$, $F(k) \approx D(k)$, and $F(k) < D(k)$, respectively, where $D(k)$ is the dissipation spectrum.  We observe Kolmogorov's spectrum ($k^{-5/3}$) for wavenumbers where  $F(k) \approx D(k)$ or $\Pi_u(k) \approx \mathrm{const}$.  Thus, a detailed investigation of the kinetic energy flux, the energy supply due to buoyancy, and the dissipation spectrum provide valuable inputs that rule out {\em BO} scaling for RBC, contrary to the predictions of Procaccia and Zeitak~\cite{Procaccia:PRL1989}, L'vov~\cite{Lvov:PRL1991},  L'vov and Falkovich~\cite{Lvov:PD1992}, and Rubinstein~\cite{Rubinstein:NASA1994}.   The entropy flux for RBC is constant in the inertial range, but the entropy spectrum exhibit dual branch, whose origin is related to the thermal boundary layer.  

In summary, stably stratified flows exhibit {\em BO} scaling in buoyancy dominated regime.  Turbulent convection however exhibits Kolmogorov's spectrum, rather than {\em BO} spectrum.  A recent shell model of buoyancy-driven flows~\cite{Kumar:arxiv2014_2} shows similar results.  More work, specially very large resolution simulations, are required to explore dual spectra predicted by Bolgiano and Obukhov.
\begin{acknowledgments}

Our numerical simulations were performed at Centre for Development of Advanced Computing (CDAC) and IBM Blue Gene P ``Shaheen" at KAUST supercomputing laboratory, Saudi Arabia. This work was supported by a research grant SERB/F/3279/2013-14 from Science and Engineering Research Board, India. We thank Ambrish Pandey, Anindya Chatterjee, Pankaj Mishra, and Mani Chandra for valuable suggestions.
\end{acknowledgments}

\appendix
\section{Scaling of the equations}
\label{sec:appendix}
Many researchers, e.g. ~\cite{Lindborg:JFM2006,Brethouwer:JFM2007}, have nondimensionalized Eqs.~(\ref{eq:u_dim}-\ref{eq:inc_dim}) as the following. They choose the  characteristic horizontal velocity $U_\perp$ as the horizontal velocity scale,  the horizontal length $l_\perp$  and the vertical height $l_\parallel$ as the horizontal and vertical length scales respectively, $l_\perp/U_\perp$ as the time scale, $U_\perp \mathrm{Fr}_\perp^2 / \alpha$ as the vertical velocity scale where $\alpha = l_\parallel / l_\perp$ is the aspect ratio, and $U_\perp^2\rho_0/(gl_\parallel)$ as the density scale. In terms of non-dimensional variables, the  equations are
\begin{eqnarray}
D_1\bf u_{\perp} & = & -{\nabla_{\perp} \sigma} + \frac{1}{\mathrm{Re}} D_2\bf u_{\perp}, \label{eq:u_perp} \\
\mathrm{Fr}_{\perp}^{2} D_1\bf u_{\parallel} & = & -\frac{d \rho}{dz}  - \rho+ \frac{\mathrm{Fr}_{\perp}^{2}}{\mathrm{Re}} D_{2} \bf u_{\parallel}, \label{eq:u_ll} \\
D_1 \rho & = & u_{z} + \frac{1}{\mathrm{RePr}} D_2 \rho, \label{eq:rho} \\
\nabla_{\perp} \cdot \bf u_{\perp} & = & -\frac{\mathrm{Fr}_{\perp}^{2}}{\alpha^{2}}\frac{\partial {\bf u_{\parallel}}}{\partial z}  \label{eq:inc_compo},
\end{eqnarray}
where
\begin{eqnarray}
D_1 & = &\frac{\partial }{\partial t} + ({\bf u_{\perp} \cdot \nabla_{\perp}})  + \frac{\mathrm{Fr}_{\perp}^{2}}{\alpha^{2}}  {u_{z}}\frac{\partial }{\partial z} \label{eq:operator_1} \\
D_2 & = & \frac{1}{\alpha^{2}} \frac{\partial^{2}}{\partial z^{2}} + \nabla_{\perp}^{2}.  \label{eq:operator_2} 
\end{eqnarray}
Here $\mathrm{Fr}_\perp = U_\perp/(l_\perp \mathrm{N})$ is the horizontal Froude number, and $\mathrm{N} = \sqrt{(g/\rho_0)|d\bar{\rho}/dz}|$ is the Brunt-V\"{a}is\"{a}l\"{a} frequency.


\end{document}